# Student-created physics problems as an independent and equitable assessment tool

Bruce A. Schumm[1], Joy Ishii[2], and Colin G. West[3]

[1]Department of Physics, University of California, Santa Cruz CA, 95064

[2]Division of Physical and Biological Sciences, University of California, Santa Cruz CA, 95064

[3]Department of Physics, University of Colorado, Boulder CO, 80309

**ABSTRACT**

Traditional high-stakes summative assessments – timed, in-class exams accounting for a large percentage of the term's overall grade – have often received criticism from the educational community[1,2] . Such assessments tend to prize a particular "narrow bundle of skills"[3], and have been shown in some contexts to produce disparate outcomes between different demographic groups[4]. Alternative low-stakes assessments have shown potential to improve student engagement[5] and close demographic gaps[6]. In this paper, we document our use of an alternative form of assessment, in which students were asked to create and solve a problem of their own design.

**MOTIVATION**

Much work in physics education theory focuses on differences between "expert" and "novice" mindsets in problem solving [7,8]. The success of a physics course is often considered in terms of whether it has helped students to "think like a physicist"[9], and many highly-prized physics education strategies draw their inspiration in part from this aspiration. For example, expert physicists often learn by doing and in consultation with others; classroom techniques like active learning and peer instruction provide a similar opportunity for students.

In the authors' experiences, expert physicists also sometimes gain new insights while constructing novel problems for homework and exams, a task which is notoriously complex[10]. We conjecture that, with proper scaffolding, students could also benefit from the kind of thinking that goes into writing a "good" (i.e., pedagogically valuable) physics problem.

For example, when constructing a problem, one must generally determine how the necessary information can be represented. To ensure that the problem is pedagogically valuable, one must also think specifically about the web of concepts that a question will test, and consider how the problem would be "categorized" compared to others. "Sophisticated use of multiple representations," "organization of knowledge around general principles," and "ability to categorize problems" are examples of "expert like" modes of thought in physics [9, 11,12].

Moreover, prior documented instances of students creating their own problems in physics[13]

(and related fields like math[14]) have shown promise as learning tools. It seems likely such tasks could also serve as a valuable form of nontraditional, lower-stakes assessment. The authors also wondered whether such an assessment tool would have less dependence on cultural and preparatory background than traditional forms of assessment.

**PROJECT DETAILS**

The "Problem Project" was administered as an assessment in the spring quarter of 2021 in Physics 5B at the University of California at Santa Cruz. Physics 5B introduces the basic properties and descriptions of fluids, waves, and optics. This instance of the course was taught online with synchronous lectures and discussion sections. Class enrollment consisted of 59% engineering, 23% physics, 10% other STEM and 3% non-STEM intended majors, with the remaining 5% stating no major preference.

In the statement of the assignment, students were asked to create a problem that would be appropriate as a homework or exam problem in the class, in any of the subject areas covered in the course.

Students were asked to submit a written presentation of their problem with four sections. In the first two sections, students were asked to state, and then solve, the problem they devised. In the third section, students were asked why they found the problem to be interesting. In the fourth section, students were asked why they felt the problem had good pedagogical value.

The Problem Project assignment was posted in the second-to-last week of the ten-week term, and was due the day before the end of the term. The assignment was completed by 245 students.

Assessment was based on the sophistication of the problem, the accuracy of its solution, independence from previously-assigned problems, and the student's description of its pedagogical value. The grading was not blind, as the grader (Professor Bruce Schumm, an author of this paper) could see the name of the student associated with each submission. The time required to grade averaged approximately five minutes per submission. Because the Problem Project assignment was somewhat experimental, a relatively small fraction (10%) of course credit was allocated to the assignment. By comparison, a single traditional midterm and a final exam were allocated 18% and 27% of total course credit.

**OUTCOME ANALYSIS**

From a purely subjective standpoint, students' performance on the Problem Project assignment exceeded the instructor's expectations. Students' written submissions showed a high degree of engagement and considerable enthusiasm for the project. Subjectively, between two-thirds and three-fourths of the class exhibited notable creativity in the development of their problems, with a significant number of students combining creativity with an unusual degree of sophistication.

Many student submissions contained sophisticated reflection on the problem's pedagogical value, noting when a problem required use of different representations, drew on multiple overlapping concepts, or presented familiar topics from a useful new perspective. For example, one representative student noted that their problem "allows for students to realize they need to draw it out to see the information they have." Another described the last part of their three-part problem as having the "most instructional value" because it "forces the learner to go through the whole process from the previous parts in reverse."  As we note above, these various forms of representation and metacognition are regarded as indicators of "expert like" thinking in physics, yet can be difficult to demonstrate or assess with traditional exams.

Submissions at the high end of the grade distribution were particularly notable and went well beyond what would ordinarily be expected at this level. In one memorable submission, the student expressed fascination with the fact that inserting a 45° polarizer between two crossed polarizers permits light to be transmitted through the three-polarizer set, while no light would be transmitted between the two-polarizer crossed-polarizer set. Building from this, the student's problem asked what would happen in the limit of an infinite number of polarizers, each infinitesimally rotated relative to the previous polarizer, and totaling to 90° of rotation through the full stack. Using formal properties of limits, the student's solution showed that 100% of the light would be transmitted.  Approximately five percent of submissions exhibited this degree of creativity and sophistication.

Subsequent statistical analysis showed that grades for the Problem Project generally displayed significantly lower disparities when comparing between groups separated by gender or under-represented group (URG)[15] status, compared to traditional exams.

For this analysis, students were classified into four sets of mutually exclusive groups (Table 1), based on gender, URG status, eligibility for the University of California's Educational Opportunity Program (EOP)[16], or first-generation status[17]. For each set of groups, performance was compared between exams (midterm and final) and the Problem Project.

| Classification | Breakdown |
|---|---|
|  |  |
| Female/Male* | 67/171 |
| URG/Non-URG | 48/197 |
| EOP/non-EOP | 43/202 |
| First-gen/non-first-gen | 50/195 |

Table 1: the four demographic categories used in our analysis. "URG" refers to students from under-represented groups, "first-gen" refers to first generation college students, and "EOP" refers to students eligible for the Educational Opportunity Program (EOP). *Note: for the first of these classifications, data from students not reporting or reporting as non-binary were excluded from the analysis because that subsample was too small to draw statistically sound conclusions.

Table 2 shows the resulting mean performance for each of the two assessments (exams, Problem Project) for the two groups within each of the four classification schemes. Also shown is the difference between the two groups for the given assessment. For all four classification schemes, significant gaps are observed in exam performance between the groups, with male, non-URG, non-first-generation, and non-EOP students all receiving higher scores than their complementary subgroup.

However, for two of the four classification schemes (female/male and URG/non-URG), the gap between the groups is seen to be significantly reduced for the Problem Project. In both of these cases, the Problem Project performance is the same (within statistical uncertainty) between the groups. A small reduction is also observed in the size of the performance gap for first-generation students, although the effect is not statistically significant. No reduction is observed for EOP students.

| Grouping | Weighted Exam Performance | Problem Project Performance |
|---|---|---|
| | | |
| Female | $63.3 \pm 2.1$ | $71.8 \pm 2.5$ |
| Male | $71.6 \pm 1.2$ | $74.8 \pm 1.4$ |
| **Difference** | **$-8.3 \pm 2.4$** | **$-3.0 \pm 2.9$** |
| | | |
| URG | $61.9 \pm 2.0$ | $73.5 \pm 2.5$ |
| Non-URG | $70.9 \pm 1.2$ | $73.8 \pm 1.4$ |
| **Difference** | **$-9.0 \pm 2.1$** | **$-0.2 \pm 2.9$** |
| | | |
| First-Generation | $63.3 \pm 2.0$ | $69.5 \pm 2.7$ |
| Non-First-Generation | $70.6 \pm 1.2$ | $74.8 \pm 1.4$ |
| **Difference** | **$-7.3 \pm 2.3$** | **$-5.3 \pm 3.0$** |
| | | |
| EOP | $63.8 \pm 2.4$ | $68.3 \pm 3.2$ |
| Non-EOP | $70.2 \pm 1.2$ | $74.9 \pm 1.3$ |
| **Difference** | **$-6.4 \pm 2.7$** | **$-6.6 \pm 3.5$** |

Table 2: Mean exam and Problem Project performance, and error on the mean, by group. Entries show percentage of total possible credit, for the exams (average of midterm and final, weighted by percentage of total class credit) and Problem Project, and differences between groups for the exams and Problem Project. A smaller magnitude of difference for the Problem Project relative to exams lends support to the hypothesis that the Problem Project is a preferable assessment mode with respect to systemic bias.

Despite the differences in the mode of assessment, the scores on the Problem Project also showed moderate positive correlations with the students' scores on the final exam across all demographic subgroups considered above. This suggests that, even as some demographic gaps closed, the project continued to track, at least to some degree, the criteria typically captured with a summative assessment.

**SUMMARY AND CONCLUSIONS**

The "Problem Project" assessment, in which students created their own pedagogically intentioned physics problems, produced submissions which were consistently impressive for their thoughtfulness and creativity, allowing students to demonstrate competence and understanding in ways not easy to measure with traditional exams. Our statistical analysis showed that, compared to traditional exams, the assessment reduced the gap in scores that had been present between students of different genders and between URG/ non-URG students. Despite these differences, scores were still significantly correlated with the traditional exams.

Because the grading of the Problem Project was not blind, the results presented here could have been affected by the instructor's holistic perspective on the students' abilities and cannot be easily generalized; a more rigorous analysis is warranted. It would also be useful to explore whether the benefits of this nontraditional assessment type persist even if it accounted for a larger percentage of the students' grades.

We view the Problem Project as a promising supplement to traditional exams, though not a replacement, as the latter can encourage students to review the course content in a comprehensive manner and can be important and independent tools for concept assimilation.

**ACKNOWLEDGEMENTS**

The authors would like to thank Dr. Jarmila Pittermann, Professor of Ecology and Evolutionary Biology at UC Santa Cruz, for discourse and input that helped to instigate and shape the Problem Project assignment.

[14] Barwell, Richard. "Working on Word Problems." *Mathematics Teaching* 185 (2003): 6-8.

[15] Underrepresented Groups (URG) include students who self-identify as African American/Black, American Indian/Alaskan Native, and/or Hispanic/Latino.

[16] The Educational Opportunity Program provides mentorship and a variety of programs and support services to those who are first generation college students and/or from low-income and educationally disadvantaged backgrounds.

[17] A student is considered first-generation if neither parent (or the single parent in the case of single-parent households) earned a four-year college degree.